\documentclass[a4paper]{article}

\usepackage{icrc2013}
\usepackage[english]{babel}

\title{Temporal and energy behavior of cosmic ray fluxes in the periods of low solar activity}

\shorttitle{Cosmic ray fluxes in the periods of low solar activity}

\authors{
Bazilevskaya G. A.,
Kalinin M. S.,
Krainev M. B.,
Makhmutov V. S.,
Svirzhevskaya A. K.,
Svirhevsky N. S.
}

\afiliations{
 Lebedev Physical Institute of the Russian Academy of Sciences, Moscow, Russia  \\

}

\email{bazilevs@sci.lebedev.ru}

\abstract{Modulation of galactic cosmic ray intensity is governed by several mechanisms including diffusion, convection, adiabatic energy losses and drift. Relative roles of these factors change in the course of an 11-year solar cycle. That can result in the changes in the energy dependence of the 11-year cosmic ray modulation. The minimum between the solar cycles 23 and 24 was extremely deep and long-lasting which led to the record high cosmic ray fluxes low-energy particles dominating. This was a signature of unusually soft energy spectrum of the cosmic rays.  In this work we examine the energy dependence  of the 11-year modulation during the  last three solar cycles and argue that a soft  energy spectrum was observed in the minimum of each cycle however only for particles below of energy around 10 GeV.  From mid 1980s the energy dependence of cosmic rays became softer from minimum to minimum of solar activity. The work is based on the cosmic ray data of the spacecraft, balloon-borne and the ground-based observations.   }


\begin{document}
\maketitle

\section{Introduction}
Modulation of galactic cosmic rays (GCRs) by solar activity is a wonderful phenomenon which includes a lot of physical processes concerning solar physics, nuclear physics, plasma physics, geophysics etc. It is believed that diffusion, convection, adiabatic energy losses and drift lead to variability of particle fluxes in the heliosphere \cite{bib:moraal1}. An 11-year and 22-year solar cycles are the main manifestations of the GCR modulation, however, the cycles differ from each other following changes in solar activity. Therefore study of GCR modulation is very complicated problem. During periods of low solar activity the GCR fluxes are maximal. The last solar minimum between cycles 23 and 24 (hereafter cycles 23/24) was extraordinary deep and long e.g. \cite{bib:ahl}. This resulted in the record high GCR fluxes \cite{bib:svir,bib:mcd,bib:mew,bib:gus}. In our works \cite{bib:svir,bib:baz} we paid attention to unusual energy dependence of the GCR modulation in the solar activity minimum of cycles 23/24. Here we  compare this finding with GCR behavior in the previous solar minima and try to match it with conditions in the near-Earth's heliosphere.

In conventional diffusion-convection theory rigidity dependence of GCR modulation is connected to rigidity dependence of interplanetary  diffusion coefficient and caused interest of many researchers, to mention a few \cite{bib:bach,bib:loc,bib:loc1,bib:ahl1,bib:alan} and others. Bachelet et al. \cite{bib:bach} concluded that for rigidity $>$ 2 GV the solar cycle modulation function for quiet periods was constant in time from 1957 to 1965, and could not be
represented by a single power law.  Investigation of GCR modulation during 1965-2000 \cite{bib:loc1} found that in the range from 0.6 to 50 GV "the rigidity dependence  of the diffusion coefficient remainds the same throughout the decrease and recovery phases of the 22-year cosmic ray modulation, except for periods around the large transient decreases". However, study of GCR modulation during 4 solar cycles on neutron monitors (1954-1999) and  3 cycles on the  IMP spacecraft  (1972-1999) led Lockwood et al. \cite{bib:loc2} to conclusion that the GCR intensity at solar minima depended on the solar magnetic field polarity: at the neutron monitor energies it was less in the positive magnetic cycles (A$>$0) and higher in the negative magnetic cycles (A$<$0) while behavior of  GCRs below $\approx$500 MeV is opposite. It is reflected in a crossover in the GCR energy spectra related to  magnetic cycles of different polarity and indicates on the rigidity dependence of the GCR drift and, probably, diffusion processes. Crossovers were also found  from study of GCR modulation based on the balloon measurements \cite{bib:svir1}.  Ahluwalia et al. \cite{bib:ahl1} fulfilled a comprehensive  investigation of  rigidity dependence of 11-year modulation for GCRs of 1-200 GV in the solar cycles 20-23. The power-law dependence was found  with small difference from cycle  cycle. However, authors  considered the complete amplitudes of GCR intensity, i.e. decrease from minimum to maximum of solar activity. The changes  in the rigidity spectrum of GCR modulation  within the 11-year cycle were studied by Alania et al. \cite{bib:alan}. They showed that the power-law index of the GCR  modulation changed in phase with solar activity, i.e. the spectrum of the 11-year modulation was harder in minimum solar activity. This research was limited by 1967-2002 and rigidities above several GV.

In this  work we do not try to find the energy  spectrum of the 11-year GCR modulation, but only trace qualitative changes in the rigidity spectrum of the GCR  on the time scale of several months, particularly around solar minima when the GCR intensity is of the highest value.

\section{Data selection}

\begin{figure}[t]
  \centering
  \includegraphics[width=0.5\textwidth]{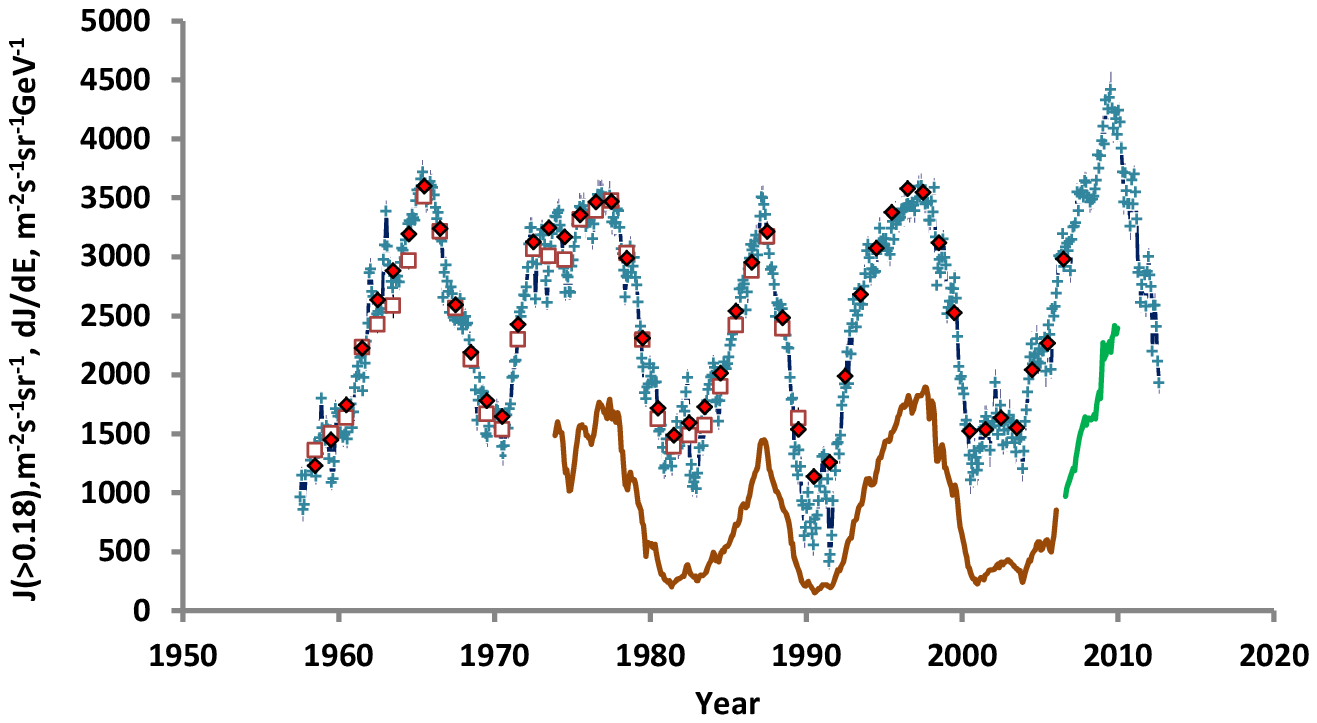}
  \caption{Galactic cosmic ray intensity vs. time. Upper curve shows monthly averaged data  J($E>$0.18 GeV)  derived from the balloon  charged particle measurements via regression method. White squares are annual values from \cite{bib:geo}. Red rhombs are annual  values from \cite{bib:preprint} (see text). Lower brown curve is plot of the IMP8 intensity of protons with E= 0.121-0.230 GeV \cite{bib:webber}. Green curve is reconstruction of intensity of protons with E= 0.10-0.22 GeV  from PAMELA results \cite{bib:adriani2}. }
  \label{fig one}
 \end{figure}

{In the last minimum of solar activity a prevailing growth of rather low energy cosmic rays was observed \cite{bib:svir,bib:baz}.  Our previous analysis included observations on balloons and on the ground-based neutron monitors. This work incorporates, in addition to balloon and ground-based observations, the results of cosmic ray measurements in space.

Monitoring of GCR is being performed during more than 6 decades - from 1950s up to present \cite{bib:simpson}.
The bulk of observational data comprise the results of the neutron monitor (NM) world-wide network \cite{bib:izmiran} with geomagnetic cutoffs $R_c$ from 0 GV to $>$10 GV. Trying to select most stable and long operating NM stations we finally have chosen Apatity ($R_c$=0.57 GV), Oulu ($R_c$=0.57 GV), Kiel ($R_c$=2.36 GV), Moscow ($R_c$=2.45 GV), Potcheftsroom ($R_c$=7.2 GV), Tsumeb ($R_c$=9.2 GV), Mexico ($R_c$=9.53 GV)  and Huancayo/Haleakala ($R_c$=12.9 GV).

 The long-term cosmic ray balloon experiment being performed at Lebedev Physical Institute (LPI) \cite{bib:stozhkov} supplements the NM data series in the lower energy range. Usually, the results of charged particle measurements in the maximum of the transition (Pfotzer) curve in the atmosphere at the polar regions  are used that  refer to effective energy about several GeV \cite{bib:moraal}. Here, we use the fluxes of GCRs on the top of the atmosphere derived from the balloon measurements at the Murmansk region ($R_c$=0.57 GV). The extrapolation of the transition curve to the top of the atmosphere gives a sum of primary GCR and albedo fluxes. A special procedure of albedo allowance was developed by A.N. Charakhchyan and T.N. Charakhchyan  \cite{bib:char}  based on the charged particle observations at  latitudes of Murmansk, Moscow and Alma-Ata. The annual values of GCR intensities with $R$ above geomagnetic cutoff over Murmansk ($E>\approx$0.18 GeV) for 1958-1988 derived from extrapolation of particle fluxes to the atmospheric boundary were published in \cite{bib:geo}. Since balloon cosmic ray observation at Alma-Ata was closed in 1993, another procedure for estimation of the primary GCR fluxes from the balloon measurement was developed based on regression between the primary GCR fluxes and the counting rate of a balloon borne detector at the Pfotzer maximum of  Murmansk region \cite{bib:preprint}. Here we have used a slightly modified procedure of regression between the primary GCR fluxes and the counting rate of a balloon borne detector at the residual pressure 8-100 g/cm$^2$. Figure \ref{fig one} demonstrates the GCR intensities with $E>$0.18 GeV as obtained with different procedures. It is seen that different approaches do not lead to significant discrepancies in the estimated intensity with $E>$0.18 GeV.
   }

 {Beginning from November 1973 to January 2006 the IMP8 data are available which refer to protons of 120-230 MeV \cite{bib:webber}. Fortunately, the PAMELA data from mid 2006 can be used to continue this data series \cite{bib:adriani2}. Actually \cite{bib:adriani2} presented the time series for protons 0.12-0.13 GeV and detailed energy spectra for periods of 2006/11/13 - 2006/12/4, 2007/11/30-2007/12/27, 2008/11/19-2008/12/15, and 2009/12/6-2010/1/1. We have taken the PAMELA  intensity of 0.12-0.22 GeV protons for  these 5 time intervals and interpolated between them according to the intensity-time profile of the PAMELA 0.12-0.13 GeV protons.

 Time  series of GCR intensity as obtained in the long-term LPI balloon experiment and in space by IMP8 and PAMELA are plotted in Fig. \ref{fig one}}.

 \begin{figure}[t]
  \centering
  \includegraphics[width=0.5\textwidth]{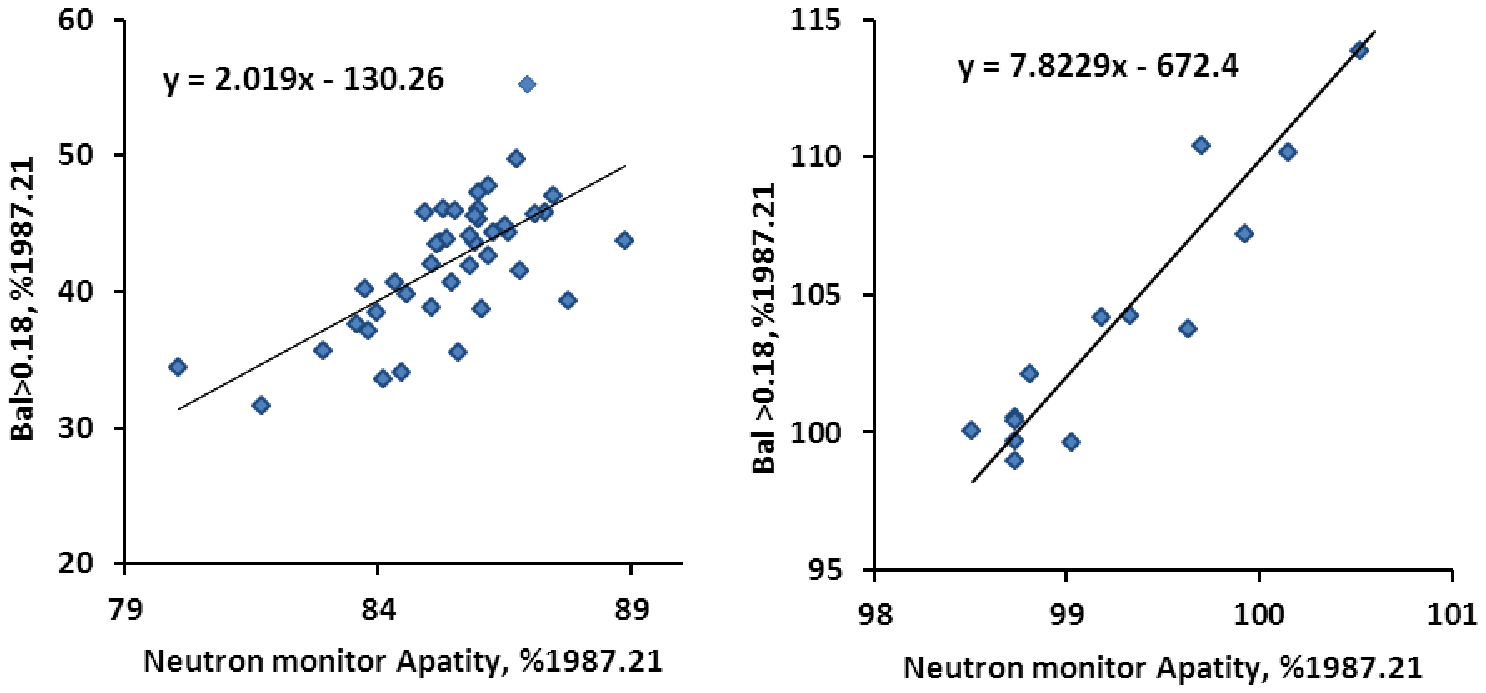}
  \caption{Scatter plots between the balloon J($E>$0.18 GeV) data ( $J>0.18$)  and the NM Apatity in June 2000-November 2003 (left panel) and November 2007-January 2009 (right panel). Data are normalized to 100\% in March 1987. The regression equations are shown.}
  \label{fig two}
 \end{figure}

\section{Energy dependence of the 11-year GCR modulation}
\begin{figure}[t]
  \centering
  \includegraphics[width=0.5\textwidth]{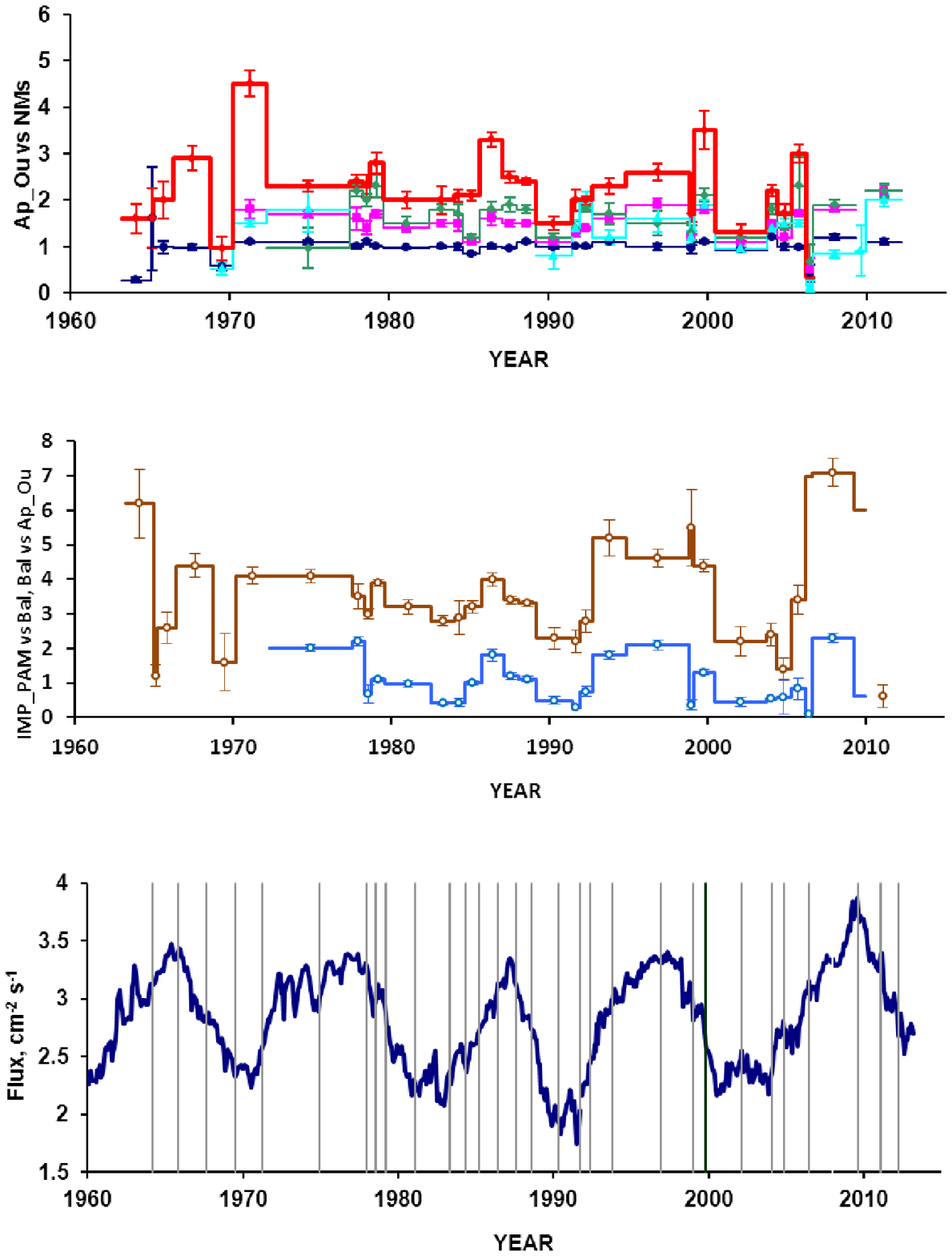}
  \caption{Bottom panel: the $\Delta T$ intervals chosen for the regression calculation on the background of cosmic ray intensity as measured on balloons in the Pfotzer maximum of Murmansk region. Vertical bars are the boundaries which isolate time intervals with more or less smoothed changes in cosmic ray fluxes. Middle panel: regression coefficients $B$ versus time. Brown line is for regression of $J>0.18$ versus Apatity-Oulu, blue one, for regression of IMP-PAMELA versus $J>0.18$. Upper panel:  coefficients $B$ for regression of Apatity-Oulu versus NMs Kiel-Moscow (blue), Potcheftsroom (magenta), Tsumeb (green), Mexico (cyan), Huancayo/Haleakala (red).  }
  \label{fig three}
 \end{figure}

{The monthly averaged data of GCR observation were treated. All the data were normalized to 100\% in March 1987. To estimate the energy dependence of the GCR variations connected to the 11-year solar activity we calculated linear regression between data series in the form $Y=A+BX$, where $Y$ and $X$ stand for the data of any two cosmic ray series. This procedure was applied to finite time intervals. It is clear that for the given time period $\Delta T$ a regression coefficient $B$ is the ratio between the changes of intensity at $Y$ and $X$ stations in this time period: $B=\Delta Y/\Delta X$. Given the stations are sensitive to different cosmic ray energy the changes in $B$  reflect the changes in the energy dependence of GCR modulation. For example, Fig.  \ref{fig two} shows scatter plots between the balloon J($E>$0.18 GeV) data (hereafter $J>0.18$)  and the NM Apatity. Left panel is for June 2000-November 2003. Change  of 20\% in the $J>0.18$ matches with 10\% at the NM, a regression coefficient $B$=2.0 which argues for a rather soft modulation spectrum. Right panel refers to November 2007-January 2009. This time change  of 15\% in the $J>0.18$ matches with 2\% at the NM, a regression coefficient $B$=7.8 which corresponds to a hard modulation spectrum.

 Actually, the temporal variations in different data sets result in high scatter of the regression coefficients when the $\Delta T$  periods are several months up to one year. To diminish the scatter we averaged the data of the NMs with close geomagnetic cutoffs and got the combined sets, namely Apatity and Oulu, Kiel and Moscow. There are few NMs with high values of cutoff no one  operating during the whole time since 1958 till present. Each of them was treated separately.

Several $\Delta T$ intervals were tested from 6  to 18 months. Eventually a variable  $\Delta T$ was applied which was chosen as periods of more or less smooth GCR intensity temporal changes. This is illustrated on the lower panel of Fig. \ref{fig three} where the monthly results of charged particles measured on balloons are plotted. Vertical bars  are boundaries of the chosen $\Delta T$ intervals. The regression coefficients $B$ were calculated always taking as $Y$ a station with response to lower energy, i.e. IMP-PAMELA versus $J>0.18$, $J>0.18$ versus Apatity-Oulu, Apatity-Oulu versus Kiel-Moscow etc. Two upper panels of Fig. \ref{fig three} present the selected results of this treatment. Middle panel gives the time dependence of regression coefficients between $Y$=IMP-PAMELA and $X=J>0.18$ and between $Y=J>0.18$ and $X$=Apatiy-Oulu. Upper panel gives the regression between $Y$=Apatity-Oulu and other NM stations taken as $X$. No clear time variations of $B$ values related to 11-year cycle can be found on this panel. From the middle panel of Fig. \ref{fig three} one can see that before mid-1980s an 11-year modulation of regression coefficients can hardly be traced. However beginning from solar cycle 21   a clear   $\approx$11-year dependence of the regression coefficients is observed with higher values during high GCR fluxes (cf bottom panel of this figure). }

\begin{figure}[h]
  \centering
  \includegraphics[width=0.5\textwidth]{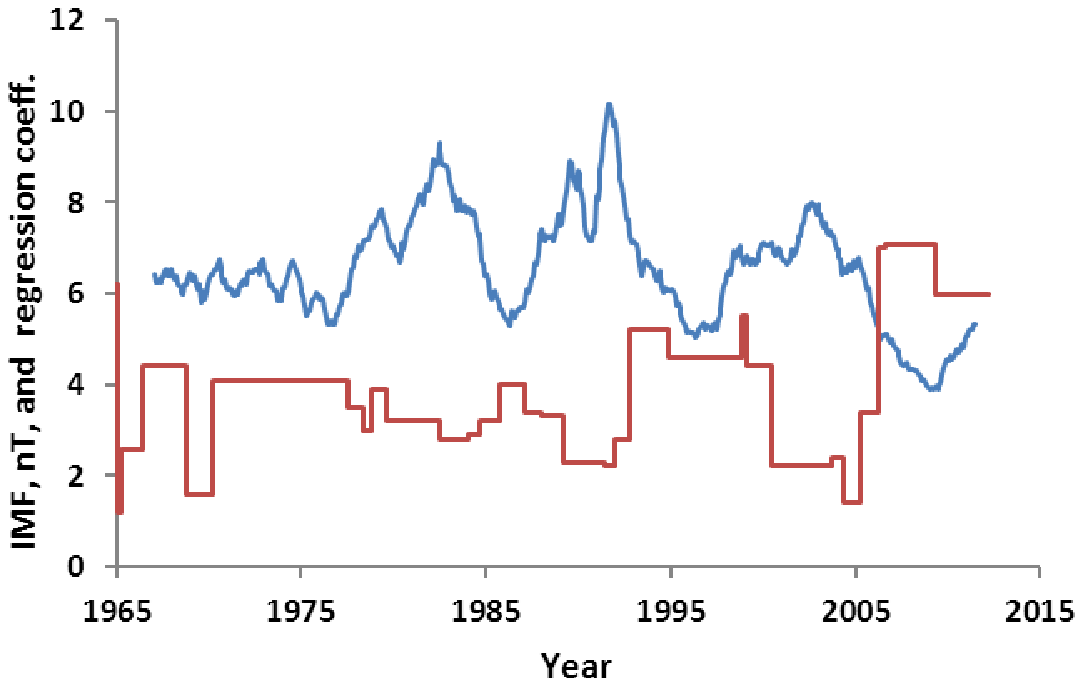}
  \caption{Time history of heliospheric magnetic field induction (blue) and rgression coefficients between  $J>0.18$ and  NM Apatity-Oulu (brown). }
  \label{fig four}
 \end{figure}

{Since low energy GCRs are subject to stronger modulation the $B$ values depend on difference in the effective energy between the data sets taken as $Y$ and $X$.  An effective energy of $J>0.18$ is estimated as $\approx<2$ GeV \cite{bib:svir1}. The effective energies of NMs are estimated by several authors \cite{bib:loc,bib:moraal,bib:usos,bib:ahl2}. The results of different approaches are not consistent with each others. In addition, the effective energies change within a solar cycle.  For the high and mid-latitude NMs various estimations are between 6 and 20 GeV, for equatorial stations, between 20 and 40 GeV.  Therefore, from Fig. \ref{fig three} we can conclude that the clear  $\approx$11-year dependence in the GCR  spectrum was observed beginning from 1980s at energies below $\approx$10 GeV: in the periods of low solar activity the  spectrum is steeper. Moreover, the spectrum became even steeper in each subsequent  minimum for solar cycles 21/22, 22/23, and 23/24. This reflected in the record high GCR intensity as observed on balloons comparing to the observed by NMs in 2009 \cite{bib:svir}.  Thus the steep GCR   spectrum in 2009 was just development of a process started in 1990s or even earlier. Surprisingly, this feature is more strongly marked in the energy range of $\approx$ 1-10 GeV ($J>0.18$ versus Apatity-Oulu) than for energies 0.1-$\approx$1 GeV (IMP-PAMELA versus $J>0.18$). This can be explained by the shift in time when the maximum of intensity was achieved according  $J>0.18$ (July 2009) and to PAMELA (December 2009). Actually, between  July and December 2009 GCR intensity was growing according to PAMELA and decreasing according to $J>0.18$. This led to decrease of the regression coefficient.

Our approach did not find any significant 11-year energy dependence of modulation for the NM energy range. Slight tendency to such dependence could be traced in the Apatity-Oulu versus Huancayo/Haleakala result (upper red line on the upper panel in Fig. \ref{fig three}) which reflects the modulation spectrum between 10 and $>40$ GeV \cite{bib:usos} but it is hardly convictive. Unfortunately, Huancayo/Haleakala station was not operative during the  solar minimum 23/24. Regression between Apatity-Oulu and other NMs virtually does not depend on the solar cycle phase.

According to the middle panel of Fig. \ref{fig three} energy spectrum of GCR  in the maximum phase of solar activity was rather hard and did not change from cycle to cycle while in the minimum phase it became softer beginning from 1980s. Our analysis is qualitative and does not allow to trace detailed transformation of the  spectrum from the hard form  to the soft and vice versa.  However, a permanent softening in the course of three solar minima could be connected to the overall change of conditions in the heliosphere. It is indicative that no clear 11-year dependence of GCR modulation can be seen before 1980 and no such dependence was observed in the heliospheric magnetic field \cite{bib:omni}. On the other hand, such a dependence has appeared after 1980 both in the heliospheric magnetic field and GCR  spectrum. During three solar cycles we observe even steeper GCR  spectrum in minima of solar activity. At this time, from 1990s, both heliospheric magnetic field induction and turbulence of magnetic field tended to weaken. Since 1990,  the solar wind density has decreased significantly which has to result in a decrease of size of heliospheric modulation region (\cite{bib:mew1,bib:kalinin,bib:krainev}). In addition, the spectrum of the magnetic inhomogeneities for the normal to the average component of the heliospheric magnetic field increased in $\approx$ 1997 rather abruptly by 20-30 \% \cite{bib:star}.

As illustration in Fig. \ref{fig four} we have plotted the time dependence of the regression coefficient between $J>0.18$ and NM  Apatity-Oulu alongside with values of heliospheric magnetic field induction. A certain negative correlation can be noticed.

Evolution of heliospheric conditions and GCR behavior during the last three solar cycles are discussed in more detail in accompanying papers  \cite{bib:kalinin,bib:krainev}. Using a rather simple model of GCR modulation and taking into account evolution of physical conditions in the heliosphere it is possible to understand and  reconstruct GCR behavior during last three cycles of solar activity both for NM energy range and lower energies.
}

 \section{Conclusion}
{Qualitative analysis of the energy dependence of the GCR modulation in the course of an 11-year solar cycle was fulfilled for low energy GCRs (below $\approx$10 GeV) and GCRs recorded by neutron monitors (above 10 GeV).  No 11-year signatures were found in the modulation energy dependence since 1960s till mid 1980s in both energy intervals. Afterwards an 11-year variation appeared in the modulation of lower energy GCRs with softer energy spectrum at minima of solar activity. Moreover, the spectrum became steeper in each subsequent  minimum for solar cycles 21/22, 22/23, and 23/24. No such behavior was observed for the GCRs of higher energy. Attempts to understand the observational data in the context of recent modulation theory are undertaken in accompanying papers \cite{bib:kalinin,bib:krainev}.

\vspace*{0.5cm}
\footnotesize{{\bf Acknowledgment:}{Authors thank colleagues maintaining operations and data processing of  neutron monitors and  spacecrafts, as well as OMNI data base.  This work was partly supported by Russian Foundation for Basic Research  (grants 11-02-00095a, 12-02-00215a, 13-02-00585a, 13-02-10006k) and by the Program "Fundamental Properties of Matter and Astrophysics" of the Presidium of the Russian Academy of Sciences.

}}

\end{document}